\documentclass[preprint,12pt]{elsarticle}

\usepackage{amssymb,color,graphicx}

\def\nat{Nat}
\def\aap{A\&A}
\def\apj{ApJ}
\def\apjl{ApJ}
\def\apjs{ApJS}

\def\mnras{MNRAS}

\def\jcap{J. Cosmol. Astropart. Phys.}
\def\prd{Phys. Rev. D}

\begin{document}

\begin{frontmatter}

\title{Neutrino emission of Fermi supernova remnants}

\author[1]{Qiang Yuan}
\author[1]{Peng-Fei Yin}
\author[1,2]{Xiao-Jun Bi}

\address[1]{Key Laboratory of Particle Astrophysics, Institute of High Energy 
Physics, Chinese Academy of Sciences, Beijing 100049, P. R. China}
\address[2]{Center for High Energy Physics, Peking University, Beijing 100871,
P.R.China}

\begin{abstract}
The Fermi $\gamma$-ray space telescope reported the observation of
several Galactic supernova remnants recently, with the
$\gamma$-ray spectra well described by hadronic $pp$ collisions.
The possible neutrino emissions from these Fermi detected
supernova remnants are discussed in this work, assuming the
hadronic origin of the $\gamma$-ray emission. The muon event rates
induced by the neutrinos from these supernova remnants on typical
km$^3$ neutrino telescopes, such as the IceCube and the KM3NeT,
are calculated. The results show that for most of these supernova
remnants the neutrino signals are too weak to be detected by the
on-going or up-coming neutrino experiment. Only for the TeV bright
sources RX J1713.7-3946 and possibly W28 the neutrino signals can
be comparable with the atmospheric background in the TeV region,
if the protons can be accelerated to very high energies. The
northern hemisphere based neutrino telescope might detect the
neutrinos from these two sources.
\end{abstract}


\end{frontmatter}

\section{Introduction}

Supernova remnants (SNR) are usually thought the most possible
candidate of the acceleration site of Galactic cosmic rays (CRs).
However, there is no firm evidence to verify such a conjecture.
High energy $\gamma$-ray observations of the sky, by e.g., the
Cherenkov telescope High Energy Stereoscopic System (H.E.S.S.) and 
the Fermi space telescope, can provide very useful information which
enables us to approach the acceleration sources of CRs. One good
example is the SNR RX J1713.7-3946. The early observation of the
very high energy (VHE) $\gamma$-ray emission from RX J1713.7-3946
by CANGAROO indicated that there might be nuclei acceleration in
this SNR \cite{2002Natur.416..823E}. The following detailed
observation made by H.E.S.S. favored a hadronic origin of the VHE
$\gamma$-ray emission according to the spectral shape
\cite{2006A&A...449..223A}. Although there are claims against the
hadronic scenario or favoring the leptonic scenario of the 
emission mechanism of SNR RX J1713.7-3946 (e.g., 
\cite{2008ApJ...683L.163L,2010MNRAS.406.1337F,2010A&A...517L...4F,
2010ApJ...712..287E}), it is still one of the most interesting 
candidates of CR nuclei acceleration sources.

After about one year's operation, Fermi collaboration reported the
observations of over $1400$ sources, within which $41$ are possible
associations of SNRs \cite{2010ApJS..188..405A}. However, it is very
likely that some of the $41$ sources are actually associated with
pulsars or pulsar wind nebulae (PWN) rather than the SNRs. According to
the morphology analysis, Fermi collaboration identified three sources
with firm association with SNRs: W44 \cite{2010Sci...327.1103A}, W51C
\cite{2009ApJ...706L...1A} and IC443 \cite{2010ApJ...712..459A}.
Further studies revealed other candidate sources of SNRs, including
Cassiopeia A \cite{2010ApJ...710L..92A}, W28 \cite{2010ApJ...718..348A},
RX J1713.7-3946 \cite{fermi:rxj1713} and W49B \cite{2010ApJ...722.1303A}.
The spectral studies of most of these sources favor hadronic origin of
the $\gamma$-ray emission (see also, \cite{2010ApJ...720...20A,
2010arXiv1007.4869O}), although for several ones the leptonic scenario
can also give an acceptable fit to the data.

It is difficult to identify the hadronic sources of CRs using
$\gamma$-rays alone. Neutrinos, if detected, can be regarded as a
definite diagnostic of the hadronic nature of the $\gamma$-ray
sources. Actually after the great progress of the discoveries of
many VHE $\gamma$-ray sources by e.g., H.E.S.S., MAGIC and Milagro,
many works studied the possible perspective of detecting neutrinos
from these sources with ongoing or upcoming neutrino detectors
(e.g.,\cite{2005APh....23..477C,2006APh....26...41C,
2006PhRvD..74f3007K,2007PhRvD..75h3001B,2007ApJ...656..870K,
2008APh....30..180G,2008A&A...484..267F,2009ApJ...695..883B,
2009A&A...495....9Y,2009APh....31..376M}). The general result is
that given the $\gamma$-rays are produced through hadronic
interactions of $pp$ collisions, some bright $\gamma$-ray sources
with hard spectra might be able to be detected with km$^3$ level
neutrino telescope like IceCube. However, the signals are usually
weak.

Since the Fermi observations of the several SNRs, especially for
those associated with molecular clouds, show strong hints of
nuclei acceleration in the SNRs, these SNRs should be prior
targets for the neutrino detection. Furthermore, the combined 
fit of the Fermi data and VHE data (if available) can help to better
determine the underlying CR spectra, and give more precise
prediction of the neutrino spectra. In this work we try to explore
the detectability of neutrinos from the Fermi detected SNRs, under
the assumption that the $\gamma$-rays are produced through $pp$
collision in the sources. The neutrino detector configuration
adopts typical km$^3$ projects, such as IceCube and KM3NeT. 

It should be noted that, however, generally we should not restrict 
the search for neutrinos on the sources which are probably of hardronic 
origin. Any sources with high enough $\gamma$-ray emission deserve 
to be paid attention to for the neutrino detection, because neutrino 
emission can independently determine the hadronic or leptonic origin 
of the sources. See e.g., \cite{2006PhRvD..74f3007K,2007PhRvD..75h3001B,
2007ApJ...656..870K,2009ApJ...695..883B} for the studies of neutrinos 
from various kinds of TeV $\gamma$-ray sources. There are also other 
studies of the neutrino emission from Fermi sources such as the blazars 
\cite{2009PhRvD..80h3008N} and the newly discovered nova 
\cite{2010arXiv1008.5193R}.

This paper is organized as follows. In Sec. 2 we first derive the
proton spectra in these SNRs to reproduce the $\gamma$-ray data
measured by Fermi and higher energy experiments. Then the neutrino
emissions of these sources are discussed in Sec. 3. We draw the
conclusion in Sec. 4.

\section{Fitting the gamma-ray spectra with hadronic model}

In this section we calculate the $\pi^0$-decay induced $\gamma$-rays
due to $pp$ collisions. We employ the parametrization given in
\cite{2006ApJ...647..692K} to calculate the differential production
spectra of secondary particles, including $\gamma$-rays and neutrinos.
The proton spectrum at the source is adjusted to reproduce the GeV-TeV
data of the SNRs. Following the ways of Fermi collaboration, we generally
adopt a broken power-law function to describe the proton spectrum.
The high energy cutoff of the proton spectrum is not well constrained
by the current TeV data. Therefore we adopt three cases for comparison:
without cutoff, exponential cutoff with $E_c=50$ TeV and $E_c=10$ TeV
respectively. However, there are two exceptions: RX J1713.7-3946 and W28.

\begin{itemize}

\item RX J1713.7-3946 --- The H.E.S.S. experiment gave very good
measurements of the TeV $\gamma$-ray spectrum of this source up to
energies exceeding $100$ TeV \cite{2007A&A...464..235A}. The
measured spectrum shows an evident curvature which disfavors a
single power-law with high significance \cite{2007A&A...464..235A}. 
We find that a power-law distribution of protons with an exponential 
cutoff can well reproduce the observational GeV-TeV data. By fitting 
the data we find the power-law index and the cutoff energy of protons
are $1.7$ and $\sim 70$ TeV respectively \cite{2010arXiv1011.0145Y}.

\item
W28 --- In the vicinity of SNR W28, H.E.S.S. observation revealed $4$
sources: HESS J1801-233 (W28 North), HESS J1800-240B (W28 South),
HESS J1800-240A and HESS J1800-240C \cite{2008A&A...481..401A}.
The Fermi observation discovered two sources coinciding with HESS J1801-233
and HESS J1800-240B. For other two sources the upper limits were given
\cite{2010ApJ...718..348A}. For these four sources we adopt a single
power-law function of the proton spectrum since it is not well constrained
by the data. As for the cutoff of proton spectrum, we also assume the
three cases described above.

\end{itemize}

\begin{figure*}
\centering
\includegraphics[width=0.45\textwidth]{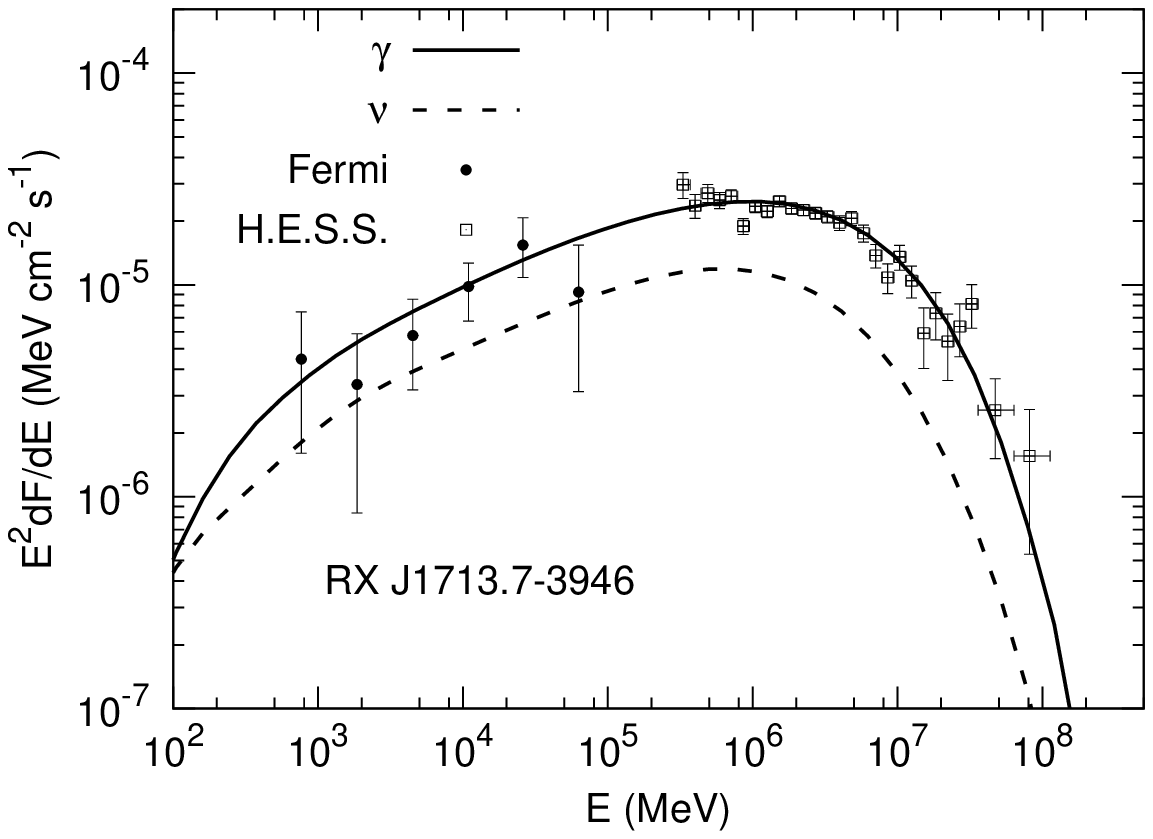}
\includegraphics[width=0.45\textwidth]{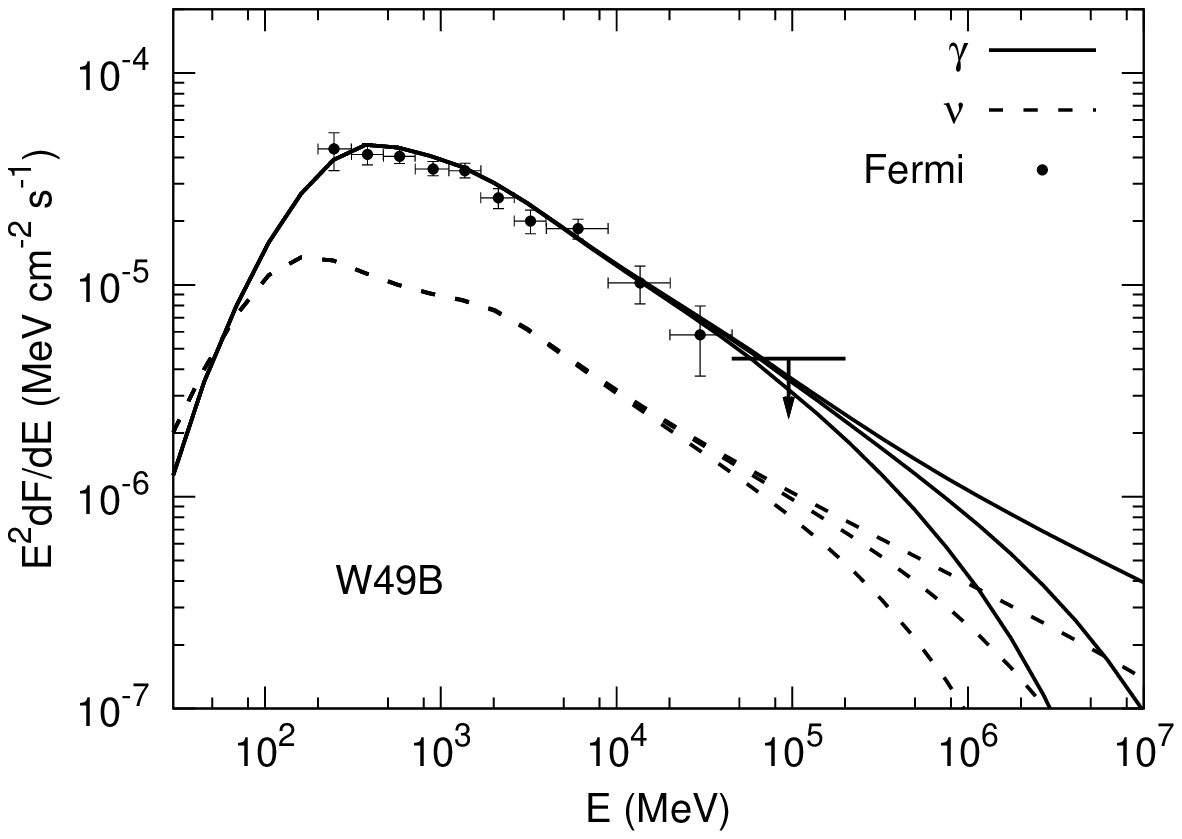}
\includegraphics[width=0.45\textwidth]{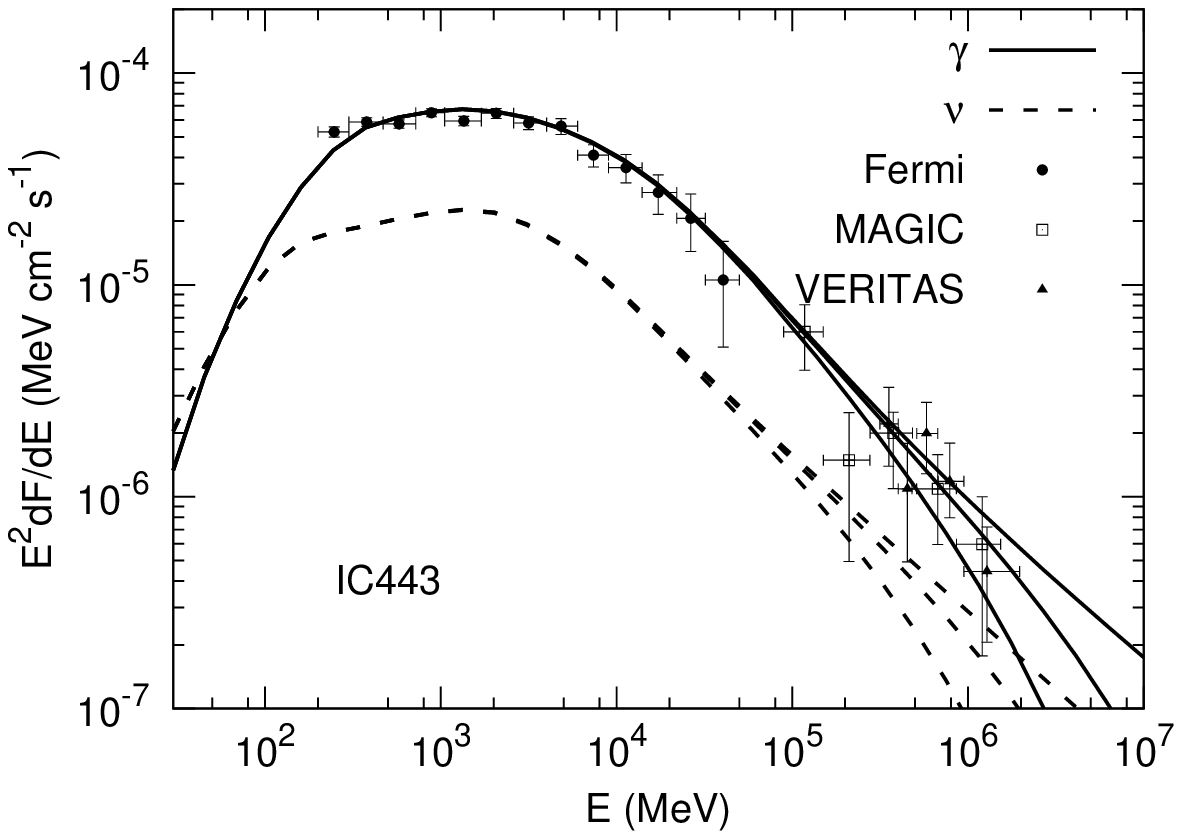}
\includegraphics[width=0.45\textwidth]{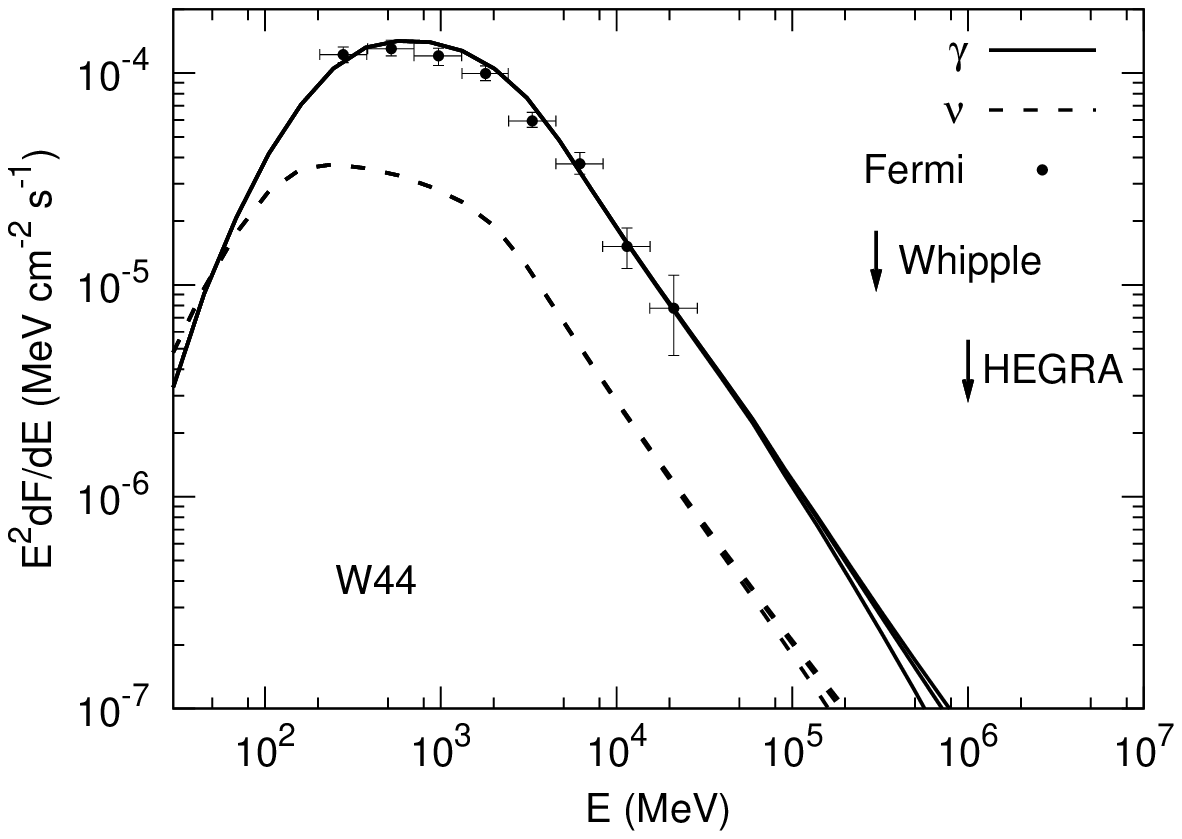}
\includegraphics[width=0.45\textwidth]{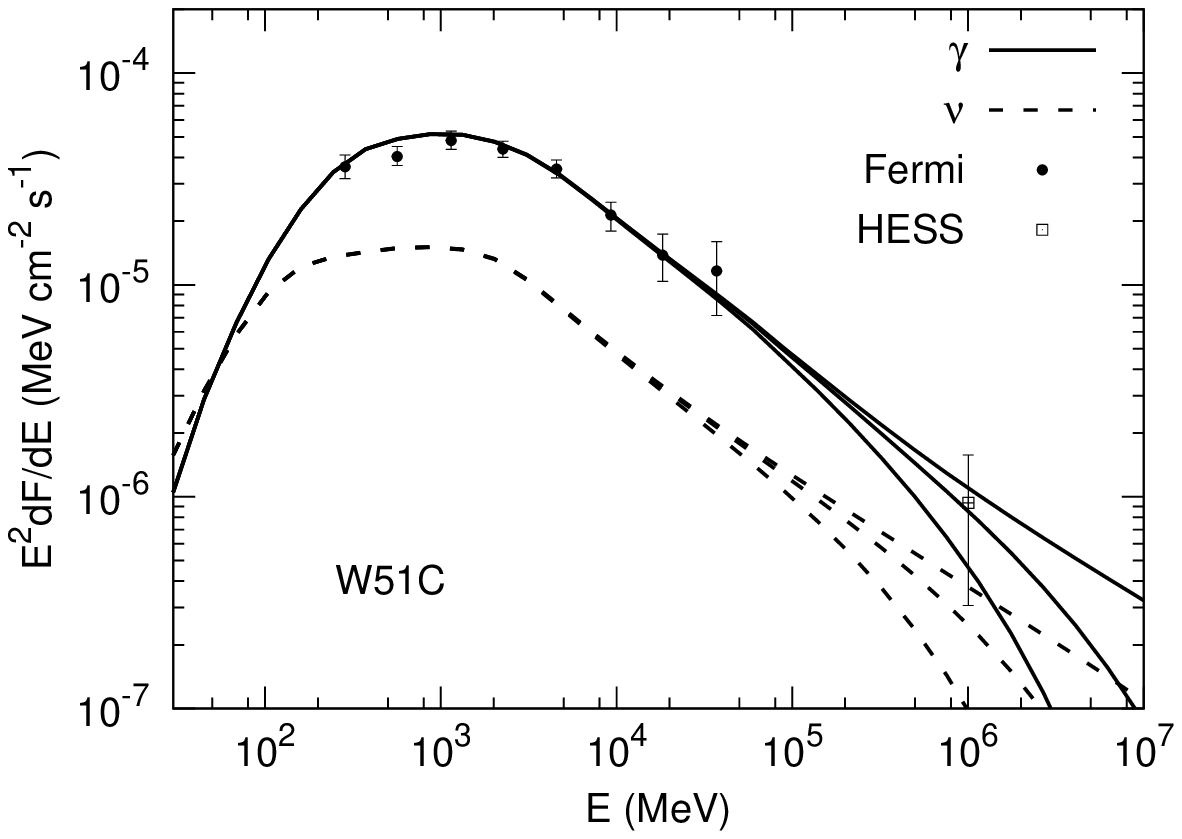}
\includegraphics[width=0.45\textwidth]{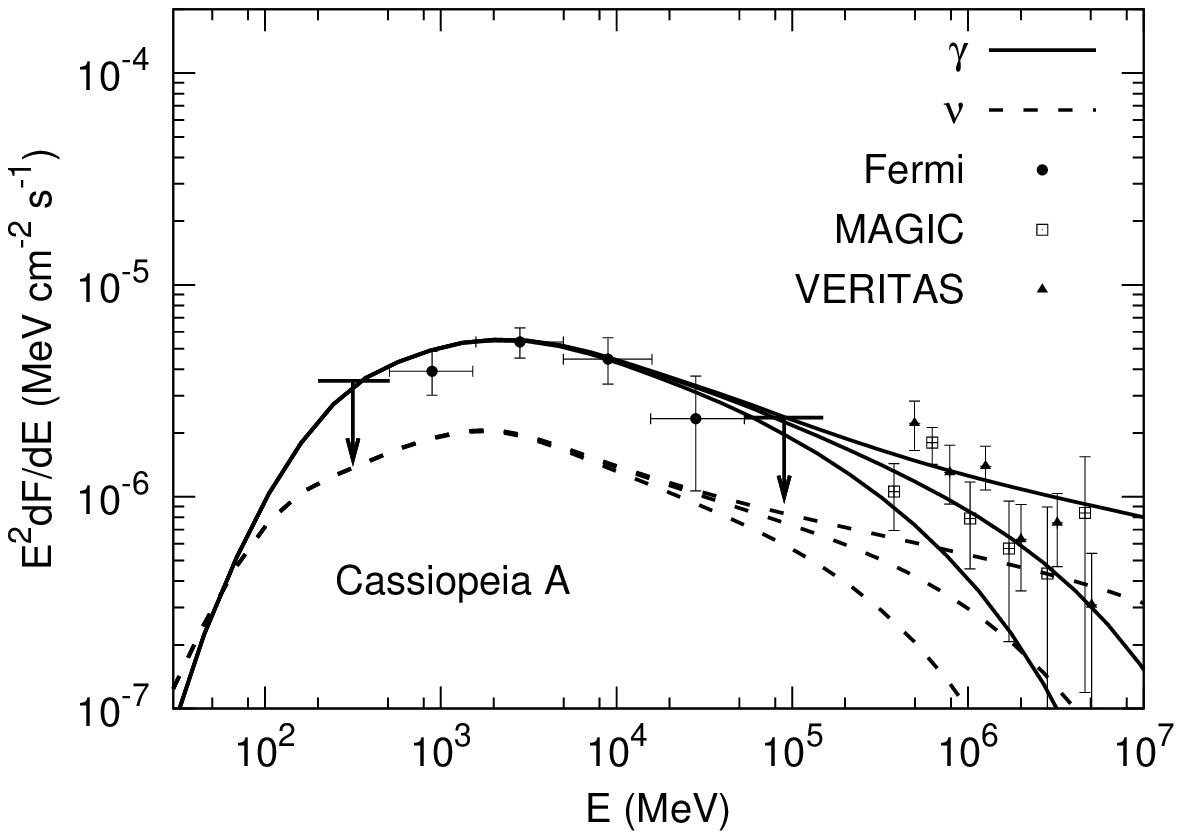}
\caption{The $\gamma$-ray spectra of seven SNRs revealed by Fermi,
together with available measurements at TeV energy range by
Cherenkov telescope. As a comparison we also show the flux of muon 
neutrinos of each source (see Sec. 3). References of the $\gamma$-ray
data are: RX J1713.7-3946, Fermi \cite{fermi:rxj1713}, H.E.S.S.
\cite{2007A&A...464..235A}; W49B, Fermi \cite{2010ApJ...722.1303A};
IC 443, Fermi \cite{2010ApJ...712..459A}, MAGIC \cite{2007ApJ...664L..87A},
VERITAS \cite{2009ApJ...698L.133A}; W44, Fermi \cite{2010Sci...327.1103A},
Whipple \cite{1998A&A...329..639B}, HEGRA \cite{2002A&A...395..803A}; W51C,
Fermi \cite{2009ApJ...706L...1A}, H.E.S.S. \cite{HESS_W51C_ICRC}; Cassiopeia A,
Fermi \cite{2010ApJ...710L..92A}, MAGIC \cite{2007A&A...474..937A},
VERITAS \cite{2008AIPC.1085..357H}.} 
\label{fig:spec}
\end{figure*}

\begin{figure*}
\centering
\includegraphics[width=0.45\textwidth]{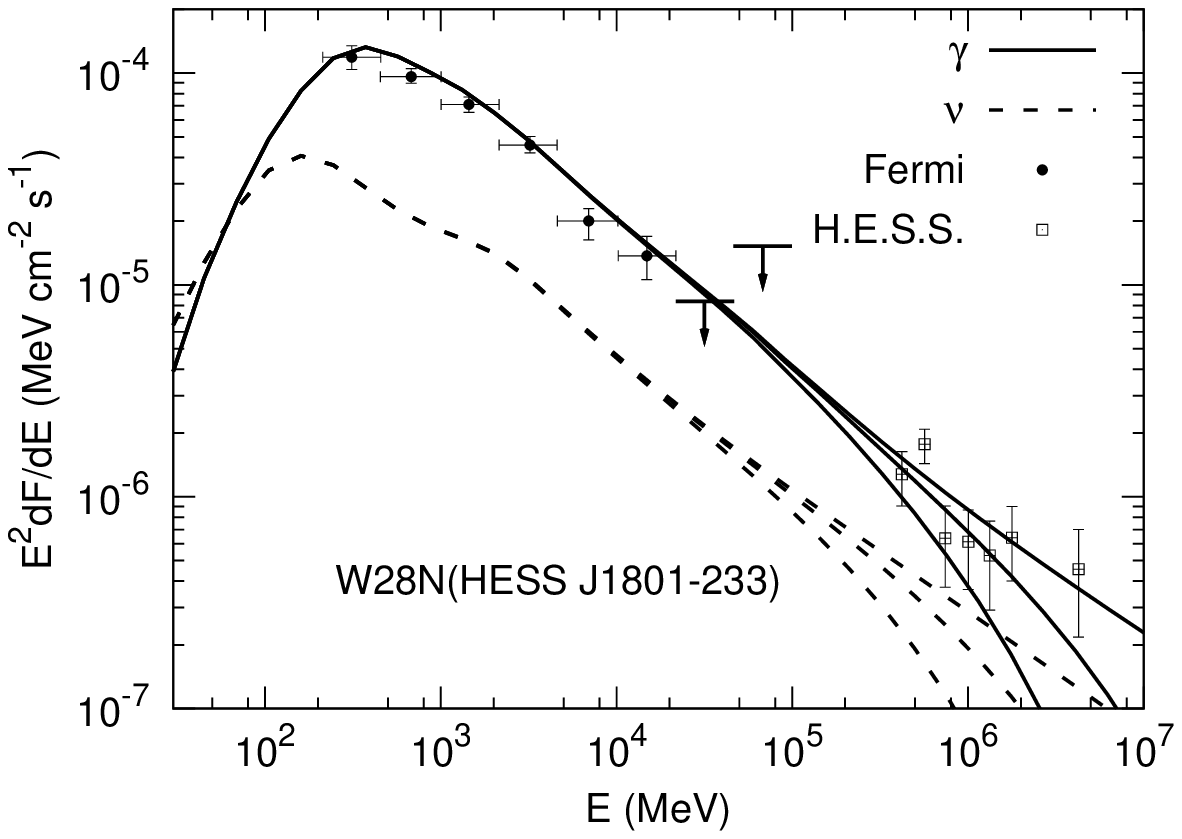}
\includegraphics[width=0.45\textwidth]{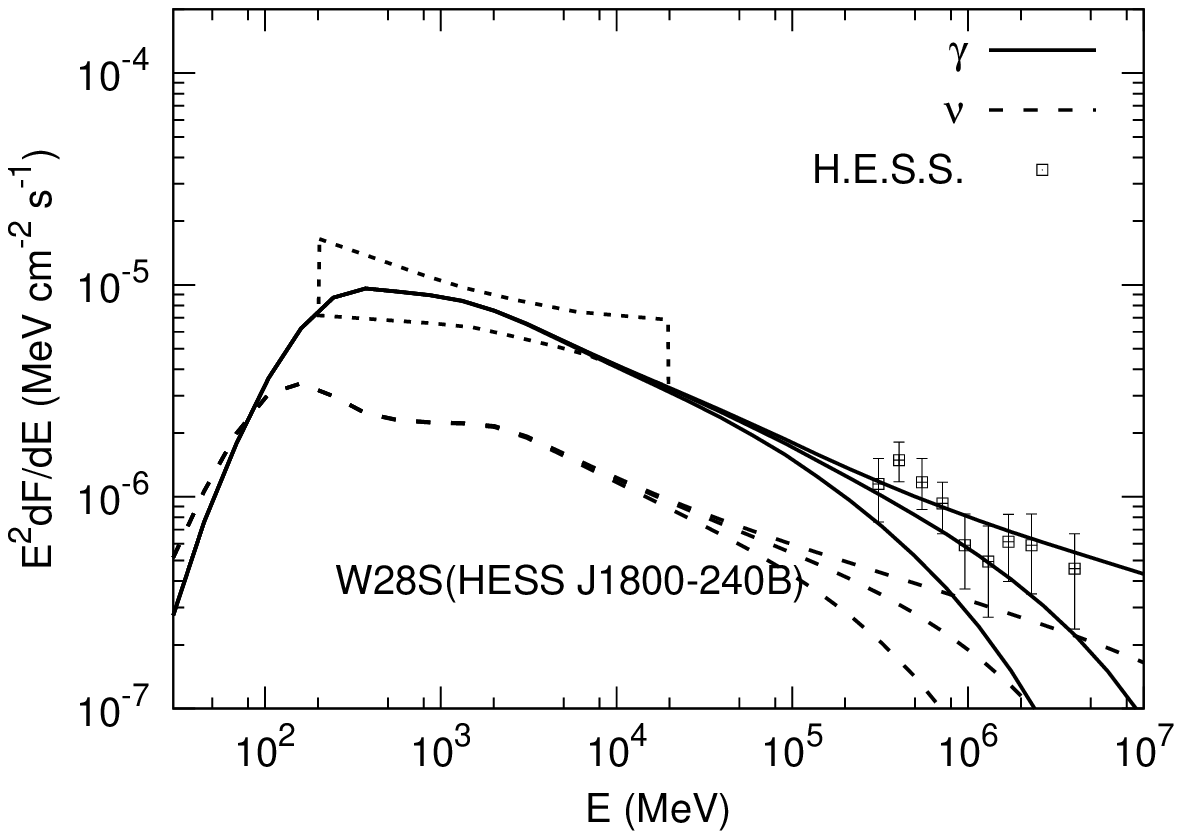}
\includegraphics[width=0.45\textwidth]{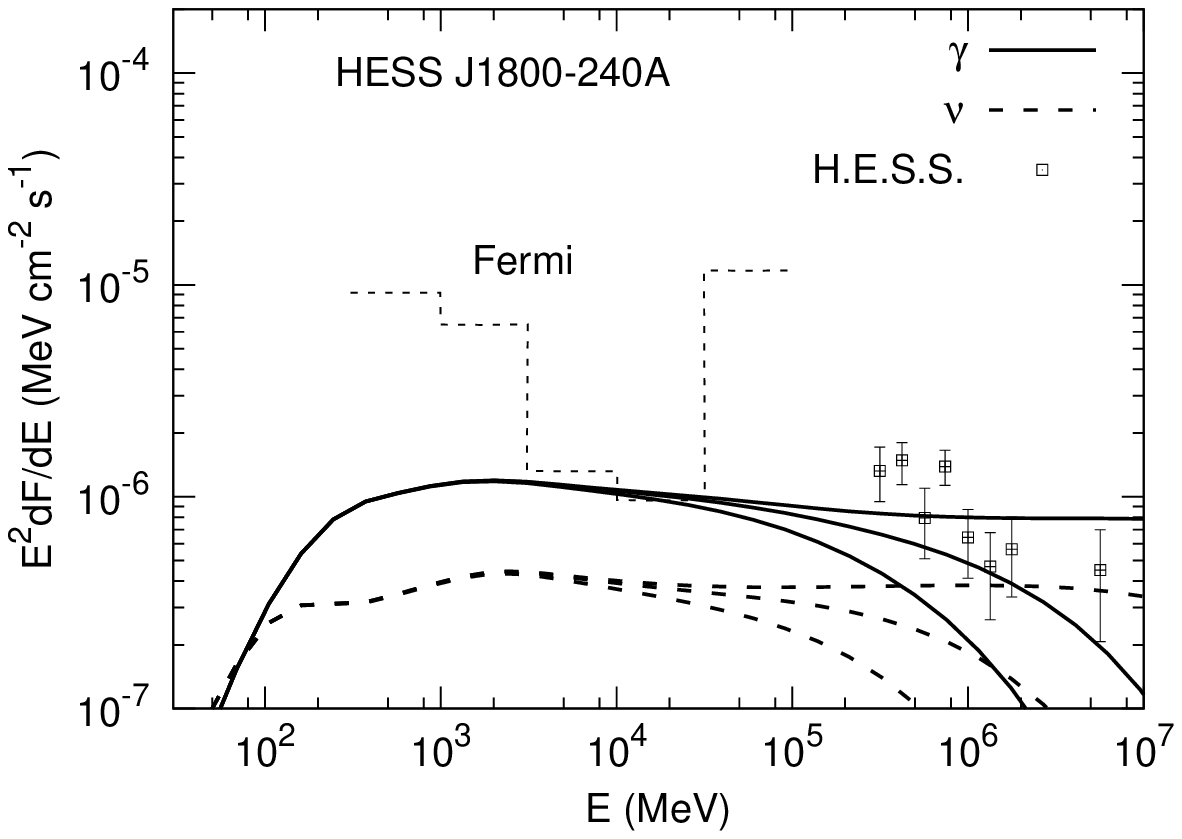}
\includegraphics[width=0.45\textwidth]{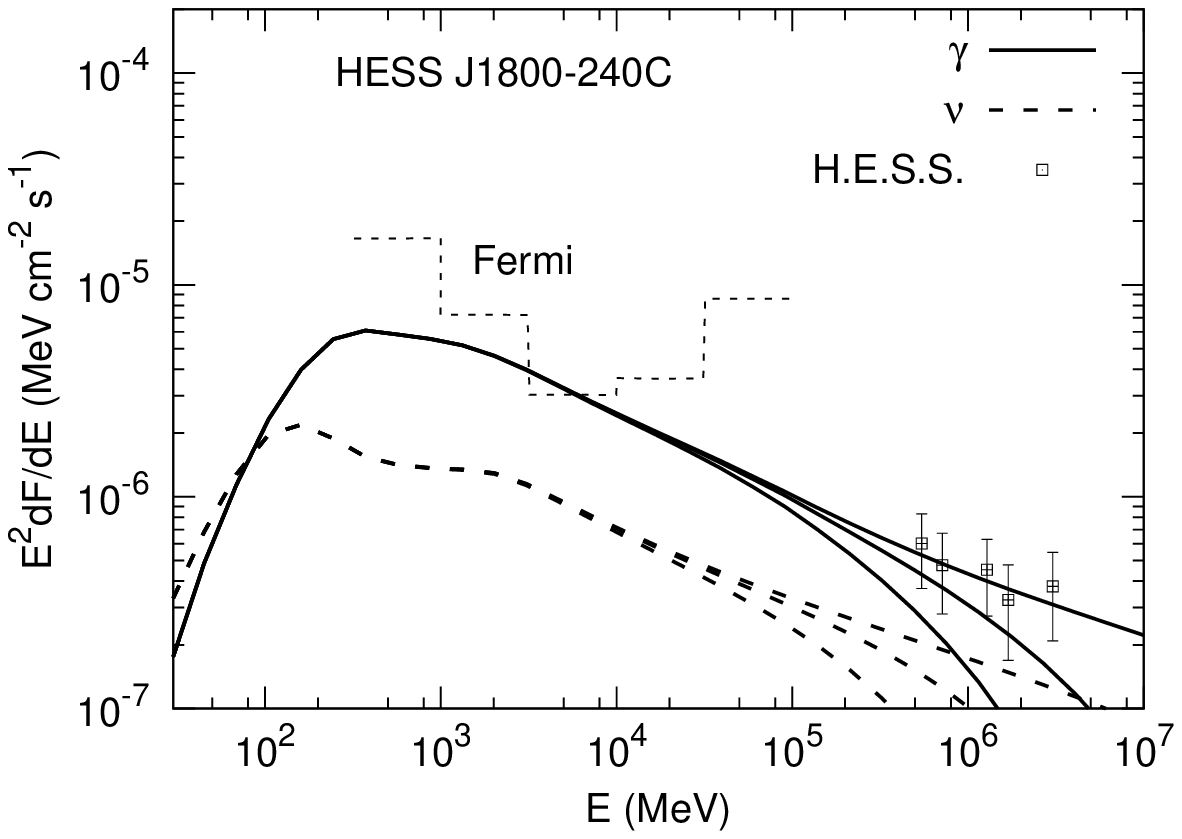}
\caption{Continuance of Figure \ref{fig:spec}. The observational data 
of Fermi are taken from \cite{2010ApJ...718..348A}, and H.E.S.S. data are
taken from \cite{2008A&A...481..401A}.}
\label{fig:spec2}
\end{figure*}

The calculated energy spectra of $\gamma$-rays from the 10
sources\footnote{For W28 there are 4 sources.} are shown in Figs.
\ref{fig:spec} and \ref{fig:spec2}. The corresponding spectral
parameters of these sources are compiled in Table
\ref{table:spec}. It is shown that generally a broken power-law
spectrum of protons can give good description to the Fermi data.
As discussed in \cite{2010arXiv1007.4869O}, this can be explained
as the escape effect of CRs from a finite-size region of the SNR.
In GALPROP, a phenomenological model for Galactic CR propagation,
the injection spectrum of CRs is also adopted as a broken
power-law function with break energy at several GeV
\cite{1998ApJ...509..212S}, which is very similar with the
results indicated by the $\gamma$-rays. This is regarded as a
support that SNRs are the sources of Galactic CRs
\cite{2010arXiv1007.4869O}. 

From the $\gamma$-ray intensity we can directly infer the neutrino
flux of these sources. The details will be given in the next
section. According to Figs. \ref{fig:spec} and \ref{fig:spec2} we
can get a rough idea that RX J1713.7-3946, Cassiopeia A and W28
should give larger neutrino signals than others.

\begin{table*}
\centering
\footnotesize{
\caption{Coordinates and proton spectral parameters of
the SNRs}
\begin{tabular}{ccccccc}
\hline \hline   & R.A. & Dec. & $\gamma_1$ & $\gamma_2$ & $E_{\rm br}$(GeV)  & $E_{\rm cut}$(TeV) \\
\hline
  RX J1713.7-3946 (G347.3-0.5) & $17^h14^m$ & $-39^\circ45'$ & $1.70$ & $1.70$ & ---  & $68$ \\
  W49B (G43.3-0.2) & $19^h11^m$ & $+09^\circ06'$ & $2.45$ & $2.55$ & $5$ & --- \\
  IC443 (G189.1+3.0) & $06^h17^m$ & $+22^\circ34'$ & $2.09$ & $2.87$ & $69$ & --- \\
  W44 (G34.7-0.4) & $18^h56^m$ & $+01^\circ22'$ & $2.00$ & $3.20$ & $10$ & --- \\
  W51C (G49.2-0.7) & $19^h24^m$ & $+14^\circ06'$ & $2.00$ & $2.65$ & $15$ & --- \\
  Cassiopeia A (G111.7-2.1) & $23^h23^m$ & $+58^\circ48'$ & $1.90$ & $2.30$ & $30$ & --- \\
  W28 North (G6.4-0.1) & $18^h00^m$ & $-23^\circ26'$ & $2.70$ & $2.70$ & --- & --- \\
  W28 South (G6.4-0.1) & $18^h00^m$ & $-23^\circ26'$ & $2.38$ & $2.38$ & --- & --- \\
  HESS J1800-240A (G6.4-0.1) & $18^h00^m$ & $-23^\circ26'$ & $2.10$ & $2.10$ & --- & --- \\
  HESS J1800-240C (G6.4-0.1) & $18^h00^m$ & $-23^\circ26'$ & $2.40$ & $2.40$ & --- & --- \\
  \hline
  \hline
\end{tabular}
\label{table:spec}}
\end{table*}

\section{Neutrino emissions and muon events}

In this section, we discuss the capability of detecting neutrino
signals from the SNRs discussed in the previous section. The SNRs
can be treated as point sources at the neutrino telescope. The
SNRs in the northern hemisphere are possibly detected by the
km$^3$ volume detector IceCube located at the South Pole. For the
sources in the southern hemisphere we discuss detectability on an
imaginary km$^3$ scale detector located in the northern hemisphere
of the Earth, such as the proposed KM3NeT.

The initial neutrino flavor ratio is approximately 
$\nu_e:\nu_\mu:\nu_\tau=1:2:0$ from the decay of $\pi^+$ and
$\pi^-$ produced by high energy $pp$ collisions\footnote{Actually
we use the parametrization given in \cite{2006ApJ...647..692K} to
calculate the initial neutrino spectrum.}. Such high energy
neutrinos arrive at the neutrino telescope after oscillations. For
the vacuum oscillation, we adopt neutrino mixing angles as
$\sin^2\theta_{12}=0.304$, $\sin^2\theta_{23}=0.5$,
$\sin^2\theta_{13}=0.01$ \cite{2008NJPh...10k3011S}, and the
neutrino flavor conversion probabilities are
$P_{\nu_e\leftrightarrow \nu_\mu}=0.22$,
$P_{\nu_\mu\leftrightarrow \nu_\mu}=P_{\nu_\mu\leftrightarrow
\nu_\tau} =0.39$ \cite{2010JCAP...04..017C}. Then the flux of
muon-neutrinos arriving at the Earth is $\Phi_{\nu_\mu}=\sum_i
\Phi^{\rm ini}_{\nu_i} P_{\nu_i\leftrightarrow \nu_\mu}\sim
\Phi^{\rm ini}_{\nu_\mu}/2$. We neglect the matter effect in the
Earth because we discuss the high energy neutrinos with $E>1$ TeV
here. In addition, if the neutrino energy is larger than 10 TeV,
the absorption effects of the Earth becomes important. Therefore,
the detected neutrino flux should be multiplied by a factor
$\exp[-\rho_N l \sigma_t(E_\nu)]$ to take into account such
absorption effects, where $\rho_N$ is the averaged nucleon
numbers, $l$ is the distance when neutrino travel through the
Earth, and $\sigma_t(E_\nu)$ is the total cross section of
neutrino-nucleon scattering. The fluxes of muon neutrinos of 
these sources are shown in Figs. \ref{fig:spec} and \ref{fig:spec2}.

The charged-current interactions between muon-neutrinos and nucleons
around/inside the detector will produce high energy muons, which can
emit Cherenkov light and then be recorded by the detector. There will 
be also an additional a few percent contribution, depending on the 
neutrino spectrum, to the muon events induced by $\nu_{\tau}$ 
\cite{2010arXiv1012.2137T}. In this work we do not include such a
contribution. The conversion probability of a muon-neutrino into a muon is
\begin{equation}
P_{\rm CC} dr\,dE_{\nu_\mu}=\left[ \frac{d\sigma_{\rm CC}^{\nu p}
(E_{\nu_{\mu}},E^0_{\mu})}{dE^0_{\mu}}\,\rho_p +(p\rightarrow n)
\right] dr\,dE_{\nu_\mu},
\end{equation}
where $\rho_p(\rho_n)$ is the number density of protons (neutrons) in
the matter. The cross sections of deep inelastic neutrino-nucleon scattering
processes are given by \cite{2007PhRvD..76i5008B}
\begin{equation}
\frac{d\sigma_{\rm CC}^{\nu\,(p,n)}(E_{\nu}, y)}{dy}
\simeq\frac{2\,m_{p,n}\,G_F^2}{\pi}\,E_{\nu}
\left(a_{\rm CC}^{\nu\,(p,n)}+b_{\rm CC}^{\nu\,(p,n)}\,(1-y)^2\right),
\end{equation}
where $y\equiv 1-E_{\ell}/E_{\nu}$, $a_{\rm CC}^{\nu\,(p,n)}=0.15, 0.25$, $b_{\rm CC}^{\nu\,(p,n)}=0.04, 0.06$ .

If the scattering between neutrino and nucleon occur inside the detector,
then the muon tracks will also start inside the detector. This is the
so-called contained event. The differential muon event rate of contained
events is given by \cite{2009PhRvD..80d3514E}
\begin{equation}
\left(\frac{d\Phi_{\mu}}{dE_{\mu}}\right)^{\rm con}=L_{\rm det}
\int_{E_{\mu}}^{\infty}dE_{\nu_{\mu}}\frac{d\Phi_{\nu_{\mu}}}
{dE_{\nu_{\mu}}}P_{CC}+(\nu\rightarrow\bar{\nu}),
\end{equation}
where $L_{\rm det}$ is the length of the detector, which is adopted to be
$1$ km.

The neutrino induced muons could also be produced in the medium around
the instrument volume. In such cases, some of the muons can propagate
into the detector and leave the tails of tracks in the detector. These
events are called through-going muons. For the high energy neutrinos,
they could produce muons which can travel a long distance, and enhance
the final muon event rate. The muons will lose energy due to ionization
and radiation when they travel in the medium. To calculate the event rate
of through-going muons, the energy losses of muons before they arrive at
the detector need to be taken into account. If we consider the average
rate of muon energy loss as $dE_\mu/dx=-\alpha-\beta E_\mu$, the distance
that a muon with energy $E^0_\mu$ can travel in matter when its energy
drops to $E_\mu$ is given by
\begin{equation}
R(E^0_\mu, E_\mu)=\frac{1}{\rho \beta} \ln \frac{\alpha+\beta E^0_\mu}
{\alpha+\beta E_\mu}.
\end{equation}
On the other hand, if the detector observes a muon with energy of
$E_\mu$, the initial muon energy at the place with a distance $r$ from
the detector could be calculated as
\begin{equation}
E^0_\mu=e^{\beta \rho r}E_\mu+(e^{\beta \rho r}-1)\frac{\alpha}{\beta}.
\end{equation}
Therefore, the event rate of through-going muons is given by
\cite{2009PhRvD..80d3514E,2010JCAP...04..017C}
\begin{equation}
\left(\frac{d\Phi_{\mu}}{dE_{\mu}}\right)^{\rm thr}=\int_{E_{\mu}}^{\infty}
dE_{\nu_{\mu}}\int_{0}^{R(E_{\nu_\mu},E_\mu)} dr \, e^{\beta \rho r}
\frac{d\Phi_{\nu_{\mu}}}{dE_{\nu_{\mu}}}P_{\rm CC}P_{\rm surv}+
(\nu\rightarrow\bar{\nu}),
\end{equation}
where the factor $dE^0_\mu/dE_\mu=e^{\beta \rho r}$ accounts for the
energy shift when the muon travel in the medium before arriving at the
detector, $P_{\rm surv}$ is the surviving probability of the muon before
decay, which roughly equals to $1$ for high energy neutrino interested
here \cite{2009PhRvD..80d3514E}.

The total muon event number at the detector is
\begin{equation}
N=\int d\Omega \int dE_\mu \frac{d\Phi_\mu}
{dE_\mu} A_{\rm det} T f(E_{\mu}),
\end{equation}
where $f(E_{\mu})$ is the energy response function with energy resolution 
width $\Delta\log_{10}E_{\mu} =0.3$, $T$ is the operation time, and 
$A_{\rm det}$ is the effective area of the detector. For the contained 
events, $A_{\rm det}$ is assumed to be $1$ km$^2$, and for the 
through-going events, $A_{\rm det}$ is the effective muon detecting 
area which is a function of muon's energy and direction. We take the
effective muon detecting area of the IceCube from 
\cite{2009APh....31..437G}. In \cite{2009APh....31..437G} the authors
also proved that such an effective muon detection area is equivalent 
to the simulated effective neutrino detection area as reported in 
\cite{2010arXiv1012.2137T,2010arXiv1007.1247H,2010JCAP...12..005A}. 
To calculate the final muon event number, we add the contained and 
through-going muons together.

The main background for high energy neutrino detection is the
atmospheric neutrinos. Here we use a parametrizations of
atmospheric neutrino flux following \cite{2009PhRvD..80d3514E},
which describes the calculated results of
\cite{2007PhRvD..75d3006H}. The high angular resolution of the
detector can be used to suppress the atmospheric neutrino
background. In this work we utilize an angle resolution of
$1^{\circ}$ (half angle of a cone).

\begin{figure*}
\centering
\includegraphics[width=0.45\textwidth]{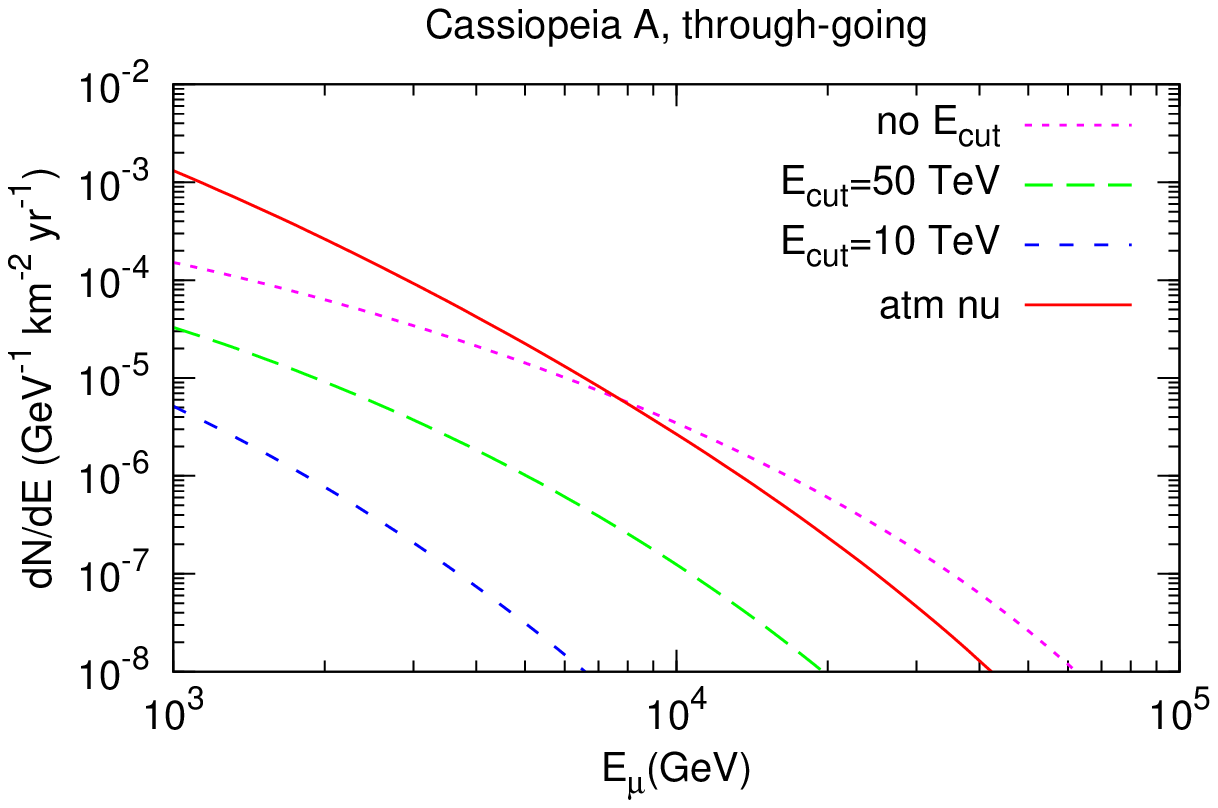}
\includegraphics[width=0.45\textwidth]{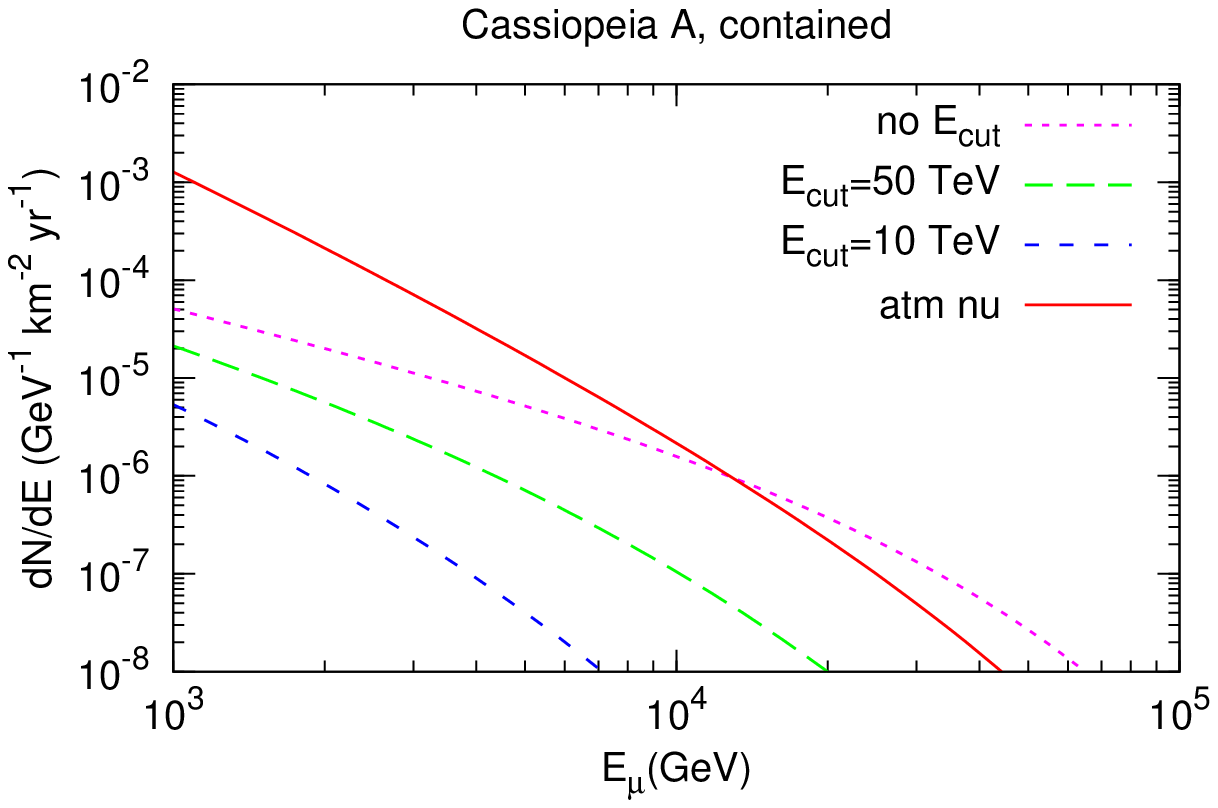}
\caption{The differential through-going (left) and contained (right) muon
rate of Cassiopiea A at IceCube. The four lines in each panel represent
muon events induced by SNR neutrinos without $E_{\rm cut}$, with
$E_{\rm cut}=50, 10$ TeV, and the atmospheric neutrinos respectively.
}
\label{fig:numu_diff}
\end{figure*}

\begin{figure*}
\centering
\includegraphics[width=0.45\textwidth]{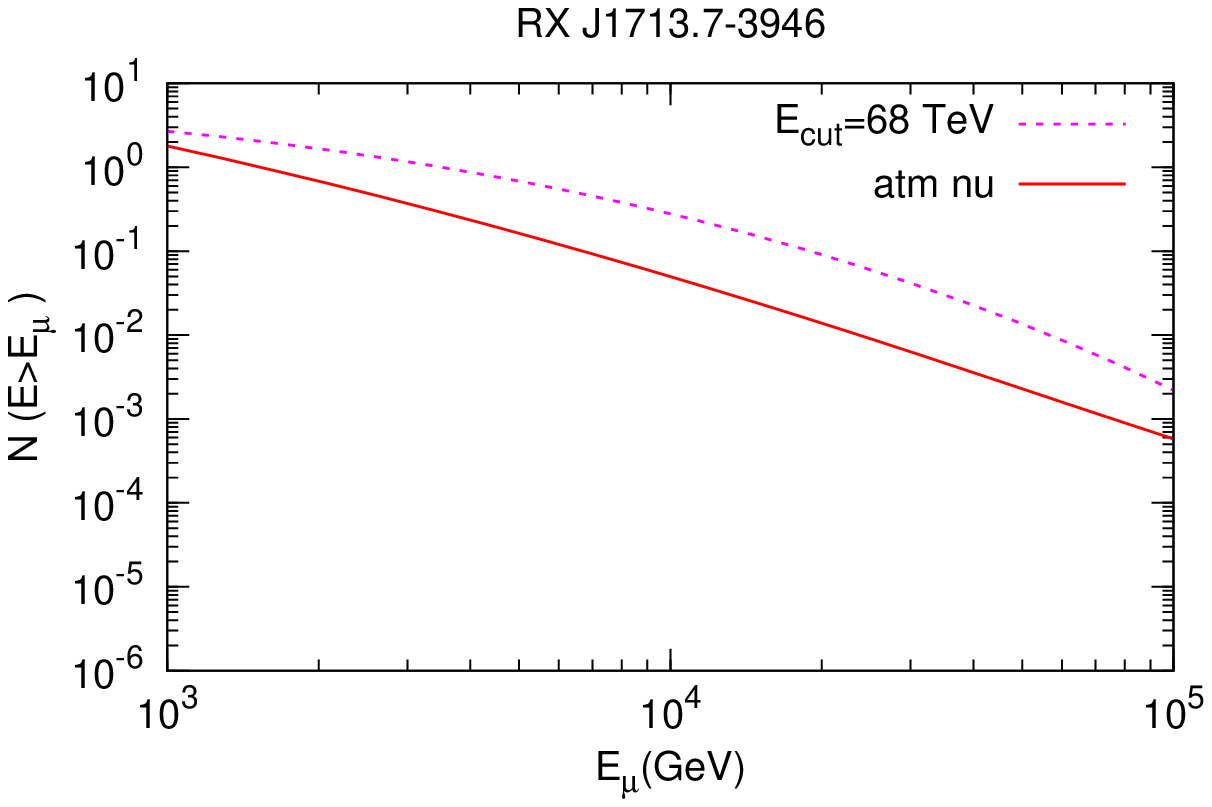}
\includegraphics[width=0.45\textwidth]{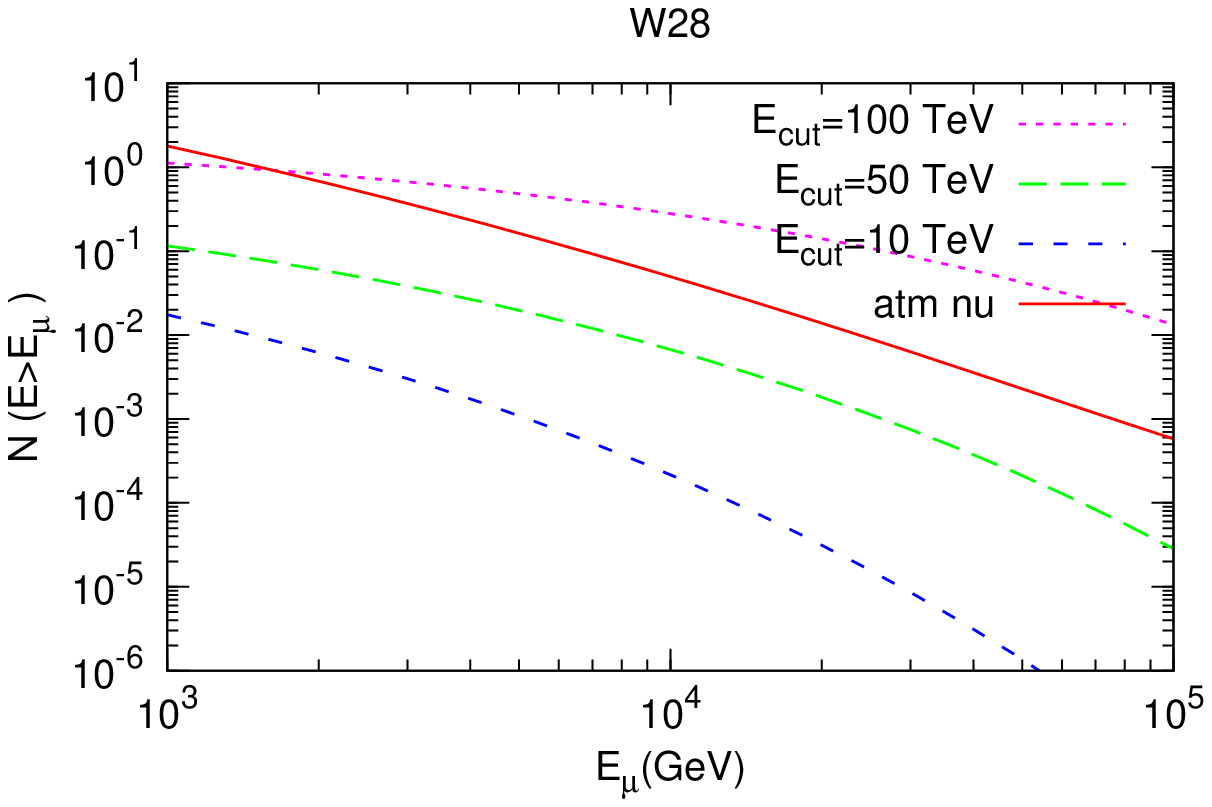}
\includegraphics[width=0.45\textwidth]{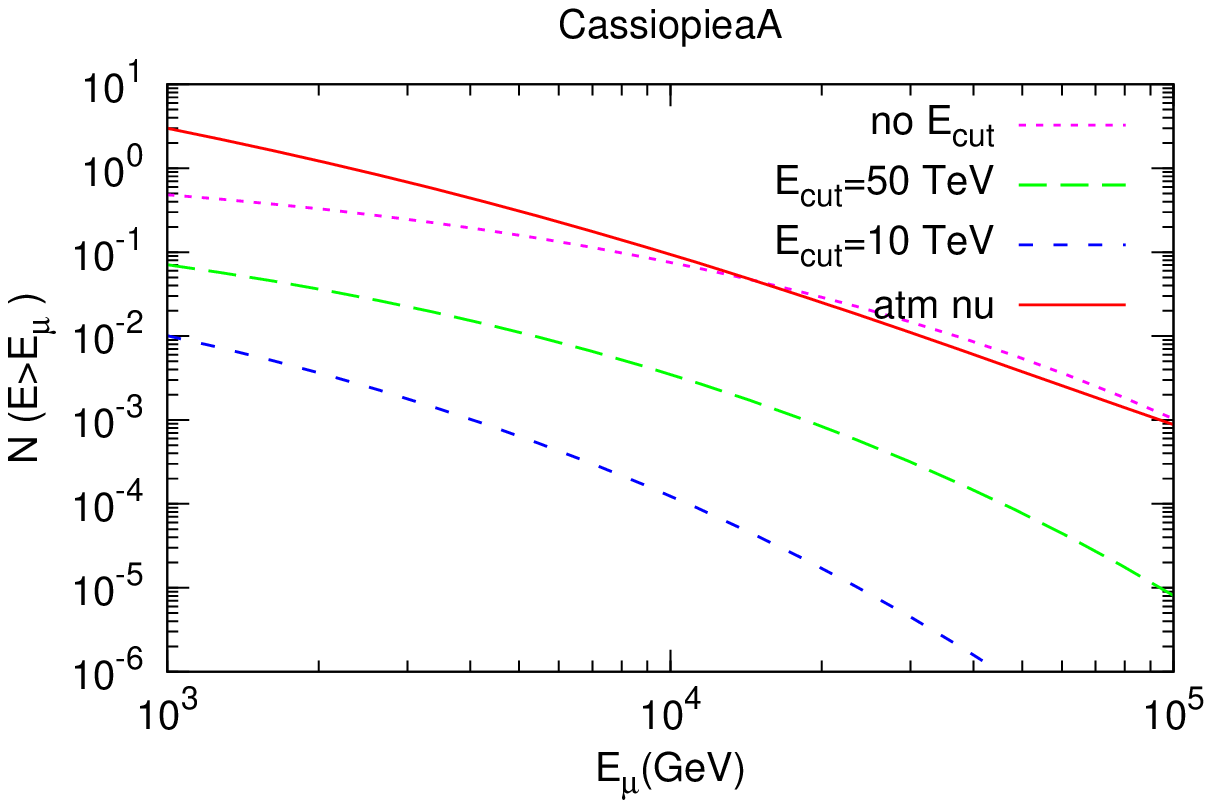}
\includegraphics[width=0.45\textwidth]{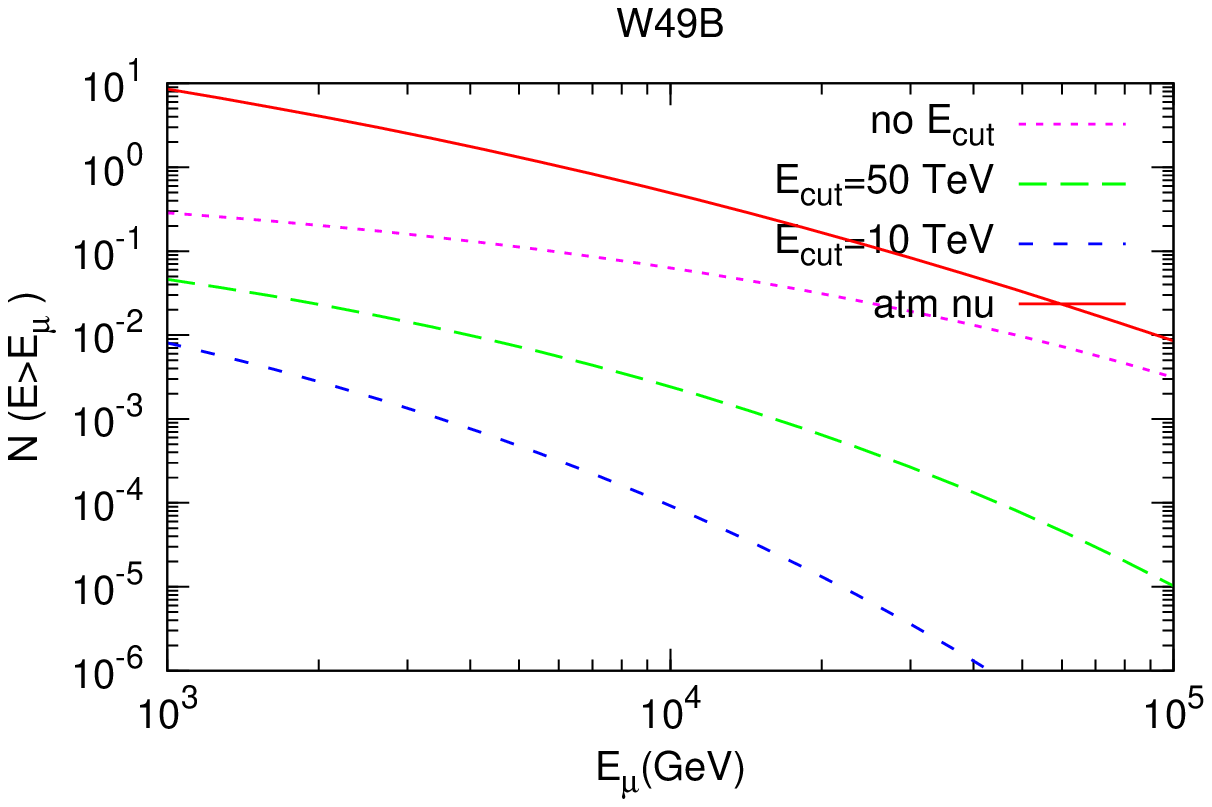}
\includegraphics[width=0.45\textwidth]{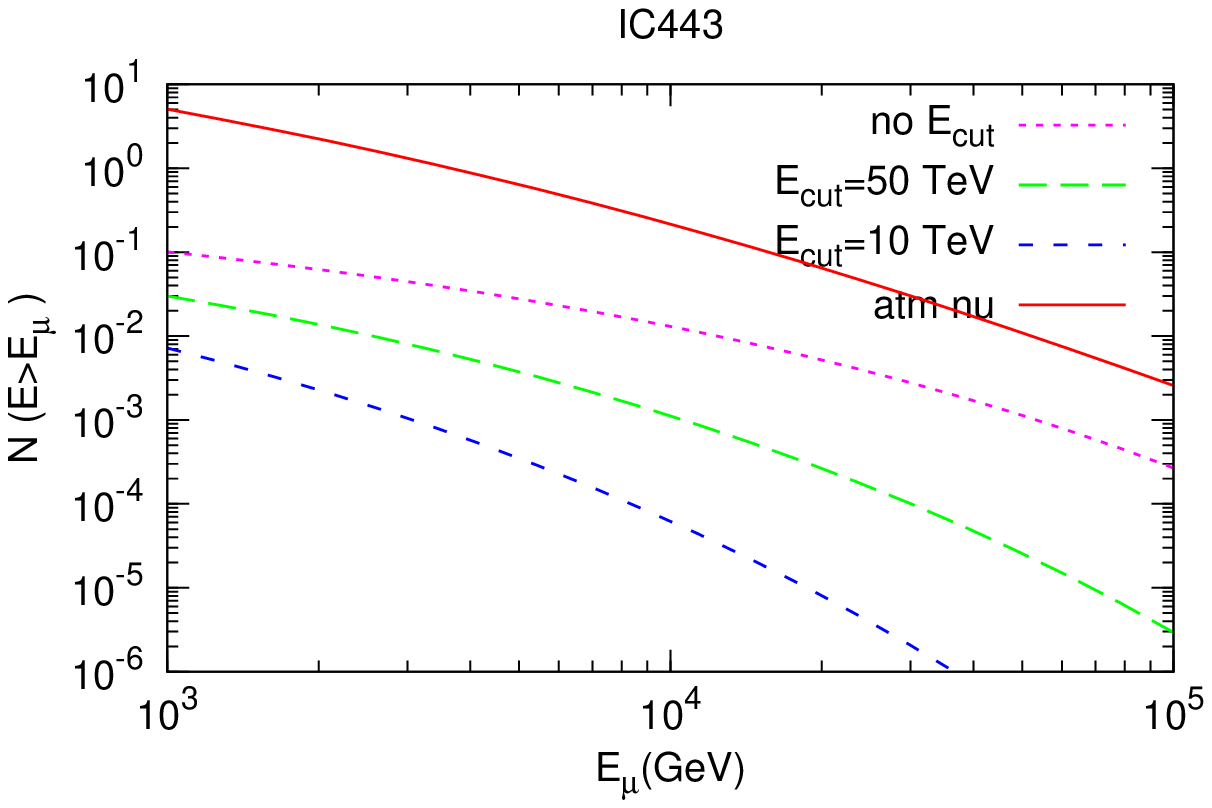}
\includegraphics[width=0.45\textwidth]{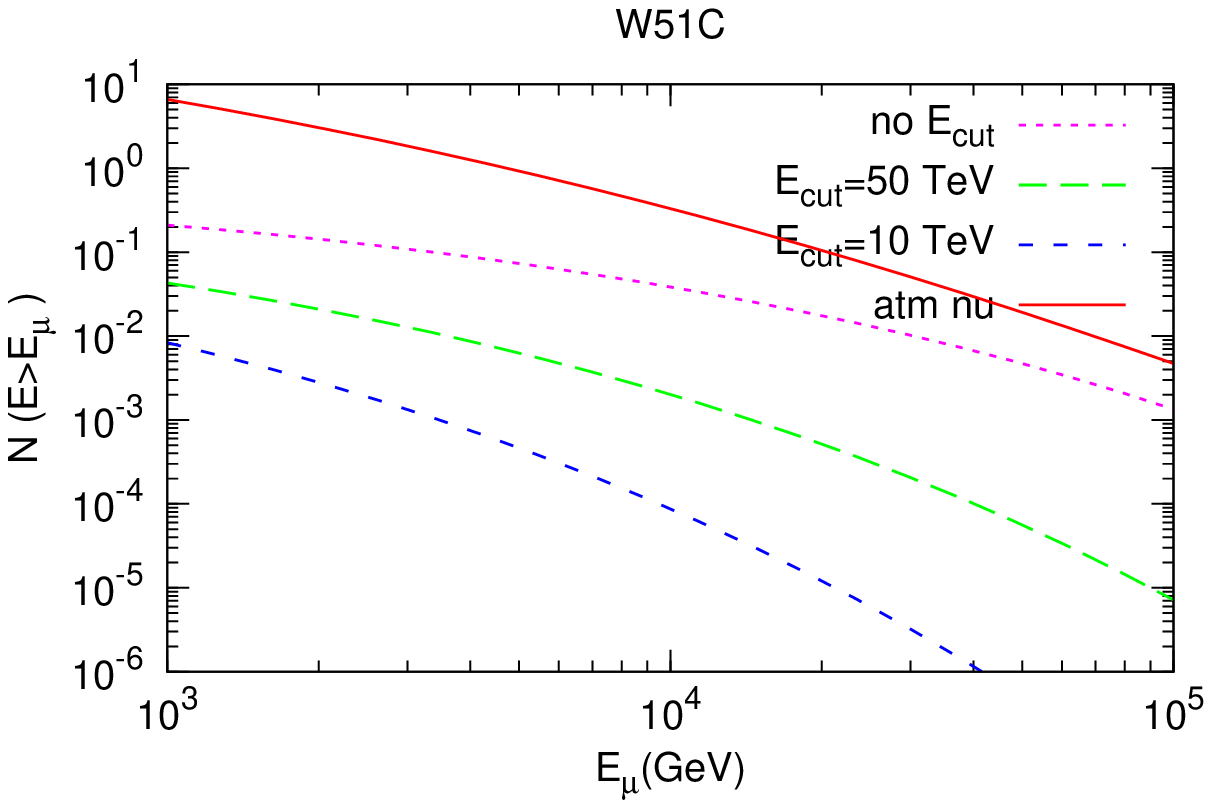}
\caption{The total muon number in one year induced by neutrino emissions
from the six Fermi SNRs. See the text for details.}
\label{fig:numu}
\end{figure*}

Compared to the SNR neutrino signals, the atmospheric neutrino background
is large but decrease quickly as $~E_\nu^{-3.7}$. Therefore we only pay
attention to muon events with energy higher than TeV. In Fig.
\ref{fig:numu_diff}, we show the differential muon rates of the 
through-going (left) and contained (right) events respectively, induced 
by neutrino emission from Cassiopeia A. According to the $\gamma$-ray
emission, we know that Cassiopeia A has a larger initial neutrino flux
with $E_\nu > 1$ TeV than other sources in the northern hemisphere.
Furthermore, the atmospheric neutrino flux from the direction of
Cassiopeia A is relatively small due to the large zenith angle.
Therefore Cassiopeia A should be the most possible candidate of
detecting neutrinos among these SNRs in the northern hemisphere,
which can be probed with the on-going IceCube detector. However,
we can see that the muon rate with energy of $O$(TeV) from
Cassiopeia A is lower by about one order of magnitude compared with
the atmospheric background, even for the case without energy cutoff of
the proton spectrum. For $E\geq 10$ TeV the signal will exceed the 
background. However, the absolute event rate is too low ($O(0.1)$) to
be detected. If there is energy cutoff of the accelerated protons
at the level of $10-100$ TeV like SNR RX J1713.7-3946, the detection
perspective of neutrinos from Cassiopeia A by IceCube would be poorer.

The total muon event numbers of six of the SNRs compiled in Table
\ref{table:spec} in one year exposure are shown in Fig.
\ref{fig:numu}. Note that the four $\gamma$-ray sources in the
vicinity of W28 are added together to be one neutrino source due
to the relatively poor angular resolution of neutrino detector.
The result of SNR W44 is not displayed either due to its extremely
soft spectrum at high energy range. The contained and
through-going muon events are added together. For the four SNRs
located in the northern hemisphere, i.e., W49B, IC443, W51C and
Cassiopeia A, the total muon event number with $E>1$ TeV is only
$O(0.1)$. Compared with the atmospheric background with the event 
rate of several, it seems to be very difficult for IceCube to discover
the neutrinos from these SNRs.

For the two SNRs located in the southern hemisphere, RX J1713.7-3946 
and W28, the situation are better. We study the detectability with a
speculated km$^3$ level neutrino telescope in the northern 
hemisphere. The actual detectability needs full Monte Carlo simulation 
based on realistic detector configuration. Here we adopt the same 
instrumental parameters of this detector as IceCube. We neglect the 
angular information of the atmospheric neutrinos, but take a directional 
averaged background. The absorption length of neutrinos when propagating
through the Earth is calculated assuming the detector is located at the
North Pole. In fact, the absorption effect is not strong. Thus, different 
neutrino traveling distances will not change the final results 
significantly. We also assume the detector only has half of the time to 
observe the southern sky per year. This is a reasonable approximation if 
the detector is not located at the North Pole. It is shown that the muon 
event number with $E>1$ TeV can reach $~3$ for RX J1713.7-3946 at such 
a detector in one year. Our result is in agreement with that derived 
in previous studies, for example, $N_{\mu}(>1\,{\rm TeV})\approx 2.8$ 
yr$^{-1}$ in \cite{2006PhRvD..74f3007K}, $\sim 2$ yr$^{-1}$ in
\cite{2007ApJ...656..870K}, $\sim 1.5$ yr$^{-1}$ in
\cite{2008A&A...484..267F}, $\sim 1.9$ yr$^{-1}$ in
\cite{2009APh....31..376M}. Compared with the background level of 
$\sim2$ yr$^{-1}$, it would be hopeful to detect the SNR neutrinos. 
This result is easy to understand because RX J1713.7-3946 is the brightest 
source in the $1-10$ TeV range among these SNRs. For W28, the number of 
muon event might reach the order of $1$ for the case without cutoff of the 
accelerated proton spectrum. For larger detectors and/or longer exposure 
time, we may still have chance to detect such neutrinos. If there
is a cutoff of the proton spectrum at energy $\sim 10-100$ TeV the
number of signal events would decrease significantly.

\section{Conclusion}

In this work we investigate the possible neutrino emissions of several
Fermi detected SNRs. Seven SNRs are compiled in this work according to
the Fermi observations, most of which are also detected by high energy
observations such as H.E.S.S., MAGIC and VERITAS. The study of the
$\gamma$-ray spectra of these sources tend to favor hadronic CRs
acceleration at the sources. Therefore the accompanied neutrino
emission will be a unique diagnostic of the nature of the radiation
of these SNRs. Assuming the $\gamma$-rays are produced through $pp$
collisions, we first determine the proton spectral parameters and intensity
normalization using the GeV band data from Fermi and the available TeV
band data from MAGIC, VERITAS or H.E.S.S.. The proton spectra are generally
adopted as a broken power-law. For RX J1713.7-3946 the TeV data measured
by H.E.S.S. are precise enough to determine the cutoff energy of protons.
However, for other sources we cannot precisely determine the high energy
behaviors using the present data. Therefore we assume $E_{\rm cut}=\infty,
\,50$ TeV and $10$ TeV for the accelerated proton spectra for comparison.
The neutrino signals are then calculated based on km$^3$ level
experiments like IceCube in the south hemisphere and KM3NeT in the north
hemisphere.

The results show that for the four SNRs located in the northern
sky, W49B, IC443, W51C and Cassiopeia A, the numbers of TeV muons
induced by the muon-neutrinos are only of the order $O(0.1)$ with
one year exposure of IceCube, whereas the atmospheric background
is larger by one or two orders of magnitude. Compared with the
large atmospheric background the detectability seems quite poor.
The two sources located in the southern sky, RX J1713.7-3946 and
W28, have larger neutrino signals than other sources. We assume an
IceCube-like detector located in the northern hemisphere of the
Earth to estimate the detectability of such northern sky sources.
We find that the number of muon events from RX J1713.7-3946 can
reach several for one year observation. The atmospheric background
in $1^{\circ}$ cone is of the same level as the signal. For W28
the signal will be slightly weaker than that of RX J1713.7-3946,
if the cutoff energy of accelerated protons is high enough. We
expect that for a long time exposure (e.g., $\sim 10$ yr) of a
km$^3$ neutrino telescope located in the northern hemisphere, it
would be possible to detect the neutrinos from the SNRs like RX
J1713.7-3946. If these neutrinos are really detected, it would be
the smoking gun of identifying the acceleration sources of the
Galactic CRs.

\section*{Acknowledgments}

This work is supported by the Natural Sciences Foundation of China 
(No. 11075169) and by the 973 project under the grant No. 2010CB833000.

\end{document}